\documentclass[dvips,prd,preprint,amssymb,nofootinbib,
letterpaper]{revtex4-1}
\usepackage{amssymb}
\usepackage[dvips]{graphicx}
\usepackage{setspace}
\begin{document}
\preprint{}

\title{LHC phenomenology of a two-Higgs-doublet neutrino mass model}

\author{Shainen M.\ Davidson}

\author{Heather E.\ Logan}
\email{logan@physics.carleton.ca}

\affiliation{Ottawa-Carleton Institute for Physics,
Carleton University, Ottawa K1S 5B6 Canada}

\date{October 5, 2010}

\begin{abstract}
We study the LHC search prospects for a model in which the neutrinos
obtain Dirac masses from couplings to a second Higgs doublet with tiny
vacuum expectation value.  The model contains a charged Higgs boson
that decays to $\ell \nu$ with branching fractions controlled by the
neutrino masses and mixing angles as measured in neutrino oscillation
experiments.  The most promising signal is electroweak production of
$H^+H^-$ pairs with decays to $\ell \ell^{\prime} p_T^{\rm miss}$,
where $\ell \ell^{\prime} = e^+e^-$, $\mu^+ \mu^-$, and $e^{\pm}
\mu^{\mp}$.  We find that a cut on the kinematic variable $M_{T2}$
eliminates most of the $t \bar t$ and $W$-pair background.  Depending
on the neutrino mass spectrum and mixing angles, a 100 (300)~GeV
charged Higgs could be discovered at the LHC with as little as 8
(24)~fb$^{-1}$ of integrated
luminosity at 14~TeV $pp$ center-of-mass energy.
\end{abstract}

\maketitle

\section{Introduction}

The Standard Model (SM) accounts for almost all experimental high
energy physics data; however, the observation of neutrino oscillations
requires that the SM be extended to include nonzero neutrino masses.
While there are many ways to expand the SM to account for
neutrino oscillations, we attempt to do so with the following goals.
First, the neutrino mass scale is significantly lower than the mass
scales of the other fermions, so we would like the model to account
for this without the addition of many tiny parameters.  Second, lepton
number violation has not yet been observed, so we would like the model
to give rise to Dirac neutrino masses, with Majorana masses forbidden.
Third, we would like the model to be testable at the CERN Large Hadron
Collider (LHC).

Most neutrino mass models give rise to Majorana masses for the SM
neutrinos, with many predicting TeV-scale new physics accessible at
the LHC.  In contrast, only a few models for Dirac neutrinos have been
proposed.  These typically involve a second Higgs doublet with very
small vacuum expectation value (vev) that couples only to the
left-handed lepton doublets and the right-handed neutrinos, resulting in
neutrino masses of the same order as the very small vev.  The original
SM-like Higgs doublet couples to all of the quarks and charged leptons
in the usual way.  Such a Yukawa coupling structure can be obtained by
imposing a global $Z_2$ symmetry, as proposed in the models of
Refs.~\cite{Ma:2000cc,Gabriel:2006ns}; however, this does not by
itself forbid neutrino Majorana mass terms, which must instead be
eliminated by imposing an additional lepton number symmetry.  The
required Yukawa coupling structure can also be obtained by imposing a
global U(1) symmetry; this idea was first proposed in
Ref.~\cite{Fayet:1974fj} as a way of ensuring the (then-assumed)
masslessness of the neutrinos in the presence of right-handed neutrino
states, and has the virtue of forbidding Majorana mass terms by
itself.

In order to generate neutrino masses, the global symmetry used to
ensure the desired Yukawa structure has to be broken.  Spontaneous
breaking leads to a very light scalar which can cause problems with
standard big-bang nucleosynthesis~\cite{Gabriel:2006ns}, as well as
having significant effects on the phenomenology of the new Higgs
particles~\cite{Gabriel:2008es}.  By instead breaking a global U(1)
symmetry explicitly, the model proposed by us in
Ref.~\cite{Davidson:2009ha} generates Dirac neutrino masses while
avoiding very light scalars.\footnote{A similar mechanism was used to
  explain the top-bottom quark mass hierarchy in
  Ref.~\cite{Hashimoto:2004xp}.}  A supersymmetric version of this
model was studied in Ref.~\cite{Marshall:2009bk}, which found
spectacular multi-lepton signals from cascade decays of the
supersymmetric partners of the new Higgs bosons and right-handed
neutrinos at the LHC.

In this paper we study the LHC detection prospects of the
non-supersymmetric model of Ref.~\cite{Davidson:2009ha}.  This model
expands the SM by adding a second Higgs doublet $\Phi_2$ with the same
electroweak quantum numbers as the SM Higgs doublet $\Phi_1$, as well
as adding three gauge-singlet right-handed Weyl spinors $\nu_{R_i}$
which will become the right-handed components of the three Dirac
neutrinos.  The model imposes a global U(1) symmetry under which the
second Higgs doublet and the right-handed neutrinos have charge $+1$,
while all the SM fields have charge zero.  This allows Yukawa
couplings of the second Higgs doublet only to the right-handed
neutrinos and the SM lepton doublet, and forbids Majorana masses for
the right-handed neutrinos.  It also tightly constrains the form of
the Higgs potential.  Breaking the U(1) symmetry explicitly using a
term $m_{12}^2 \Phi_1^\dagger \Phi_2 + {\rm h.c.}$ in the Higgs
potential yields a vev $v_2$ for the second Higgs doublet and
consequently gives the neutrinos Dirac masses proportional to $v_2$.
By requiring that $v_2 \sim \mathcal{O}({\rm eV})$, the Dirac neutrino
masses are made suitably small without requiring tiny Yukawa couplings.

The characteristic feature of the model is that the couplings of the charged
scalar pair $H^{\pm}$ and two neutral scalars $H^0$ and $A^0$ from the
second Higgs doublet to leptons and neutrinos are
controlled by the neutrino masses and mixing angles.  In this paper we
take advantage of the distinctive decay of the charged Higgs boson
$H^+$ into charged leptons and neutrinos.  We focus on electroweak
pair production of $H^+H^-$ at the LHC followed by decays to $\ell
\ell^\prime p_T^{\rm miss}$, where $\ell \ell^\prime$ can be any
combination of opposite-sign $e$, $\mu$, and $\tau$ leptons and
$p_T^{\rm miss}$ denotes missing transverse momentum (carried away by
the neutrinos).  Because $\tau$ leptons are more difficult to
reconstruct experimentally, we concentrate on the final states with
$\ell \ell^{\prime} = e^+ e^-$, $\mu^+ \mu^-$, and $e^\pm \mu ^\mp$.
The major backgrounds are diboson production ($W^+W^-$, $ZZ$, and $Z
\gamma$) and top quark pair production with both tops decaying
leptonically.

To determine whether the $H^+H^-$ signal will be detectable at the
LHC, we generated signal and background events using MadGraph/MadEvent
version 4~\cite{Alwall:2007st} assuming 14~TeV $pp$ center-of-mass
energy.  We present results both at parton level, and after
hadronization with PYTHIA~\cite{PYTHIA} and fast detector simulation
with PGS~\cite{PGS}.  With appropriate cuts, we find that a 5$\sigma$
discovery can be achieved with luminosity in the range 8--75~fb$^{-1}$
for $M_{H^+} = 100$~GeV, depending on
the neutrino mixing parameters.  For $M_{H^+} = 300$~GeV a 5$\sigma$
discovery can be made with luminosity in the range 24--460~fb$^{-1}$.
The higher luminosity requirements
occur when the neutrino parameters are such that $H^+$ decays mostly
to $\tau\nu$, leading to final states not considered in our analysis.
We find that the kinematic variable $M_{T2}$ is very effective at
separating the signal from the $t \bar t$ and $WW$ backgrounds for
charged Higgs masses above the $W$ mass, and also provides sensitivity
to the charged Higgs mass.\footnote{While we have not made a detailed study of charged Higgs detection
prospects at 7~TeV $pp$ centre-of-mass energy, we note that the cross
section for the most dangerous $WW$ background is about 2.5 times
smaller at 7~TeV.  However, the signal cross section is also about 2.5
(4.5) times smaller at this energy for $M_{H^+} = 100$ (300)~GeV.
Furthermore, the LHC is anticipated to collect only about 1~fb$^{-1}$
of integrated luminosity at 7~TeV.  We thus expect detection or even
exclusion of the process considered here to be unfeasible in the
current 7~TeV LHC run.}

This paper is organized as follows.  In the next section we review the
model and present the charged Higgs decay branching ratios.  In
Sec.~\ref{sec:SB} we describe the signal and background processes, our
event generation procedure and selection cuts, and the resulting
signal significance.  In Sec.~\ref{sec:conclusions} we summarize our
conclusions.

\section{The model}

As outlined in the introduction, we start with the field content of
the SM and add to it a new scalar SU(2)$_L$ doublet $\Phi_2$ (the SM
Higgs is denoted $\Phi_1$) and three right-handed gauge singlets
$\nu_{R_i}$ (these are the right-handed neutrinos).  We impose a U(1)
symmetry under which $\Phi_2$ and the three $\nu_{R_i}$ have charge
+1 and all the other fields are uncharged, which leads to the
Yukawa coupling structure~\cite{Davidson:2009ha}
\begin{equation}
  \mathcal{L}_{\rm Yuk} = - y^d_{ij} \bar d_{R_i} \Phi_1^{\dagger} Q_{L_j} 
    - y^u_{ij} \bar u_{R_i} \tilde \Phi_1^{\dagger} Q_{L_j} 
    - y^{\ell}_{ij} \bar e_{R_i} \Phi_1^{\dagger} L_{L_j} 
    - y^{\nu}_{ij} \bar \nu_{R_i} \tilde \Phi_2^{\dagger} L_{L_j}
    + {\rm h.c.}
    \label{eq:smyuk}
\end{equation}
Here $\tilde \Phi_i \equiv i \sigma_2 \Phi_i^*$ is the conjugate Higgs
doublet and $y^f_{ij}$ are the 3$\times$3 Yukawa matrices for fermion
species $f$.

The Higgs doublets can be written explicitly as
\begin{eqnarray}
  \Phi_i=\left( \begin{array} {c} \phi^+_i \\ 
    (v_i + \phi^{0,r}_i + i \phi^{0,i}_i)/\sqrt{2} \end{array} \right),
\end{eqnarray}
where $v_1$ will be generated by the usual spontaneous symmetry
breaking mechanism of the SM and $v_2$ will be generated by the
explicit breaking of the global U(1), described below.  Inserting
these expressions for $\Phi_i$ into Eq.~(\ref{eq:smyuk}), we obtain
the fermion masses and couplings to scalars.  In particular, the
fourth term in Eq.~(\ref{eq:smyuk}) gives rise to the neutrino mass
matrix and interactions:
\begin{eqnarray}
  \mathcal{L}_{\rm Yuk} 
  &\supset& - \frac{y^{\nu}_{ij} v_2}{\sqrt 2} \bar \nu_{R_i} \nu_{L_j} 
  - \frac{y^{\nu}_{ij}}{\sqrt 2} \phi_2^{0,r} \bar \nu_{R_i} \nu_{L_j} 
  - i \frac{y^{\nu}_{ij}}{\sqrt 2} \phi_2^{0,i} \bar \nu_{R_i} \nu_{L_j} 
  + y^{\nu}_{ij} \phi_2^+ \bar \nu_{R_i} \ell_{L_j} + {\rm h.c.}
  \label{eq:nuyuk}
\end{eqnarray}
After diagonalizing the mass matrix in the first term, the neutrino
mass eigenvalues are given by $m_{\nu_i} = y^{\nu}_i v_2/\sqrt{2}$,
where $y^{\nu}_i$ are the eigenvalues of $y^{\nu}_{ij}$.  In this way,
the small masses of the three neutrinos can be traced to the small
value of $v_2$.

We obtain the vevs of the scalar doublets from the Higgs potential as
follows.  The most general gauge-invariant scalar potential for two
Higgs doublets is (see, e.g., Ref.~\cite{HHG}),
\begin{eqnarray}
  V &=& m_{11}^2 \Phi_1^{\dagger} \Phi_1 + m_{22}^2 \Phi_2^{\dagger} \Phi_2
  - \left[ m_{12}^2 \Phi_1^{\dagger} \Phi_2 + {\rm h.c.} \right] 
  \nonumber \\
  && + \frac{1}{2} \lambda_1 \left( \Phi_1^{\dagger} \Phi_1 \right)^2
  + \frac{1}{2} \lambda_2 \left( \Phi_2^{\dagger} \Phi_2 \right)^2
  + \lambda_3 \left( \Phi_1^{\dagger} \Phi_1 \right) 
  \left( \Phi_2^{\dagger} \Phi_2 \right) 
  + \lambda_4 \left( \Phi_1^{\dagger} \Phi_2 \right)
  \left( \Phi_2^{\dagger} \Phi_1 \right) \nonumber \\
  && + \left\{ \frac{1}{2} \lambda_5 \left( \Phi_1^{\dagger} \Phi_2 \right)^2
  + \left[ \lambda_6 \Phi_1^{\dagger} \Phi_1 
      + \lambda_7 \Phi_2^{\dagger} \Phi_2 \right] \Phi_1^{\dagger} \Phi_2
    + {\rm h.c.} \right\}.
\end{eqnarray}
Imposing the global U(1) symmetry eliminates $m_{12}^2$, $\lambda_5$,
$\lambda_6$, and $\lambda_7$.  The global U(1) symmetry is broken
explicitly by reintroducing a small value for $m_{12}^2$.  This leaves
the Higgs potential~\cite{Davidson:2009ha},\footnote{Note that
  using a $Z_2$ symmetry instead of the global U(1) would allow a
  nonzero $\lambda_5$ term.}
\begin{eqnarray}
  V &=& m_{11}^2 \Phi_1^{\dagger} \Phi_1 + m_{22}^2 \Phi_2^{\dagger} \Phi_2
  - \left[ m_{12}^2 \Phi_1^{\dagger} \Phi_2 + {\rm h.c.} \right] 
  \nonumber \\
  && + \frac{1}{2} \lambda_1 \left( \Phi_1^{\dagger} \Phi_1 \right)^2
  + \frac{1}{2} \lambda_2 \left( \Phi_2^{\dagger} \Phi_2 \right)^2
  + \lambda_3 \left( \Phi_1^{\dagger} \Phi_1 \right) 
  \left( \Phi_2^{\dagger} \Phi_2 \right) 
  + \lambda_4 \left( \Phi_1^{\dagger} \Phi_2 \right)
  \left( \Phi_2^{\dagger} \Phi_1 \right).
\end{eqnarray}
Stability of the potential at large field values requires
$\lambda_1,\lambda_2>0$, $\lambda_3 > -\sqrt{\lambda_1 \lambda_2}$,
and $\lambda_4 > -\sqrt{\lambda_1 \lambda_2} - \lambda_3$.  We want
$v_1$ to arise through the usual spontaneous symmetry breaking
mechanism, which is achieved when $m_{11}^2<0$.  We do not want the
global U(1) to also be broken spontaneously, as that will create a
very light pseudo-Nambu-Goldstone boson, which is incompatible with
standard big-bang nucleosynthesis; thus we require that the curvature
of the potential in the $v_2$ direction at zero $\Phi_2$ field value
be positive, i.e., $m_{22}^2 + (\lambda_3+\lambda_4) v_1^2/2 > 0$.

To find the values of the vevs in terms of the parameters of the Higgs
potential, we apply the minimization conditions,
\begin{eqnarray}
  \left. \frac{\partial V}{\partial |\Phi_1|}\right|_{\rm min}
  &=& m_{11}^2 v_1 - m_{12}^2 v_2 
  + \frac{1}{2} \lambda_1 v_1^3 
  + \frac{1}{2} ( \lambda_3 + \lambda_4 ) v_1 v_2^2 = 0
  \nonumber \\
  \left. \frac{\partial V}{\partial |\Phi_2|}\right|_{\rm min} 
  &=& m_{22}^2 v_2 - m_{12}^2 v_1
  + \frac{1}{2} \lambda_2 v_2^3
  + \frac{1}{2} ( \lambda_3 + \lambda_4 ) v_1^2 v_2 = 0.
\end{eqnarray}
Since we will require $m_{12}^2 \ll v_1^2$, we can ignore $m_{12}^2$ and
$v_2$ when finding the value of $v_1$.  This yields
\begin{equation}
  v_1^2 = \frac{-2 m_{11}^2}{\lambda_1}.
\end{equation}
For $v_2$, we need to consider $m_{12}^2$, although again we may
ignore higher order terms in $m_{12}^2/v_1^2$; this yields
\begin{equation}
  v_2 = \frac{m_{12}^2 v_1}
  {m_{22}^2 + \frac{1}{2} (\lambda_3 + \lambda_4) v_1^2}.
  \label{eq:v2}
\end{equation}
We will choose parameters so that $v_1 \simeq 246$~GeV and $v_2
\sim$~eV.  This requires $m_{12}^2 \sim ({\rm MeV})^2$.  We note that
because $m_{12}^2$ is the only source of breaking of the global U(1)
symmetry, its size is technically natural; i.e., radiative corrections
to $m_{12}^2$ are proportional to $m_{12}^2$ itself and are only
logarithmically sensitive to the high-scale
cut-off~\cite{Davidson:2009ha}.

The mass eigenstates of the charged and CP-odd neutral scalars are given by 
\begin{eqnarray}
  G^+ &=& \phi_1^+ \sin\beta + \phi_2^+ \cos\beta \simeq \phi_1^+ 
  \nonumber \\
  H^+ &=& \phi_1^+ \cos\beta - \phi_2^+ \sin\beta \simeq - \phi_2^+
   \nonumber \\
  G^0 &=& \phi_1^{0,i} \sin\beta + \phi_2^{0,i} \cos\beta \simeq \phi_1^{0,i}
  \nonumber \\
  A^0 &=& \phi_1^{0,i} \cos\beta - \phi_2^{0,i} \sin\beta 
  \simeq - \phi_2^{0,i},
\end{eqnarray}
where we define $\tan\beta \equiv v_1/v_2 \sim 10^{11}$.  $G^+$ and
$G^0$ are the Goldstone bosons, which do not appear as physical
particles in the unitarity gauge.  $H^+$ and $A^0$ are the physical
charged and CP-odd neutral Higgs states and are almost entirely
contained in $\Phi_2$.  Neglecting contributions of order $m_{12}^2$
and $v_2^2$, the masses of $H^+$ and $A^0$ are~\cite{Davidson:2009ha}
\begin{eqnarray}
  M_{H^+}^2 &=& m_{22}^2 + \frac{1}{2} \lambda_3 v_1^2 
  \nonumber \\
  M_A^2 &=& m_{22}^2 + \frac{1}{2} (\lambda_3 + \lambda_4) v_1^2
  = M_{H^+}^2 + \frac{1}{2} \lambda_4 v_1^2.
\end{eqnarray}
The mass matrix for the CP-even neutral states is almost diagonal,
yielding only very tiny mixing of order $v_2/v_1$.  Ignoring the
mixing, the eigenstates are $h^0 \simeq \phi^{0,r}_1$ (SM-like) and
$H^0 \sim \phi^{0,r}_2$, with masses~\cite{Davidson:2009ha}
\begin{eqnarray}
  M_h^2 &=& m_{11}^2 + \frac{3}{2} \lambda_1 v_1^2 
  = \lambda_1 v_1^2 \nonumber \\
  M_H^2 &=& m_{22}^2 + \frac{1}{2} (\lambda_3 + \lambda_4) v_1^2
  = M_A^2.
\end{eqnarray}

After diagonalizing the neutrino mass matrix, Eq.~(\ref{eq:nuyuk})
yields the following couplings to the new physical Higgs states:
\begin{equation}
\mathcal{L}_{\rm Yuk} \supset -\frac{m_{\nu_i}}{v_2} H^0 \bar \nu_i \nu_i
  + i \frac{m_{\nu_i}}{v_2} A^0 \bar \nu_i \gamma_5 \nu_i
  - \frac{\sqrt{2} m_{\nu_i}}{v_2} [U_{\ell i}^* H^+ \bar \nu_i P_L e_{\ell}
    + {\rm h.c.}],
\end{equation}
where $U_{\ell i}$ is the Pontecorvo-Maki-Nakagawa-Sakata (PMNS)
matrix, defined according to $\nu_{\ell} = \sum_i U_{\ell i} \nu_i$,
where $\nu_{\ell}$ are the neutrino flavor eigenstates.

The PMNS matrix can be parameterized in terms of three mixing angles
$\theta_{ij}$ (with $ij = 12$, 23, and 13) and a phase $\delta$ according
to (see, e.g., Ref.~\cite{Fogli:2005cq}),
\begin{equation}
  U_{\ell i} = \left( \begin{array}{ccc}
    c_{12} c_{13} & s_{12} c_{13} & s_{13} e^{-i \delta} \\
    -s_{12} c_{23} - c_{12} s_{23} s_{13} e^{i \delta} 
    & c_{12} c_{23} - s_{12} s_{23} s_{13} e^{i \delta} 
    & s_{23} c_{13} \\
    s_{12} s_{23} - c_{12} c_{23} s_{13} e^{i \delta}
    & -c_{12} s_{23} - s_{12} c_{23} s_{13} e^{i \delta}
    & c_{23} c_{13} \end{array} \right),
\end{equation}
where $c_{ij} \equiv \cos \theta_{ij}$ and $s_{ij} \equiv \sin
\theta_{ij}$.  The 2$\sigma$ experimentally-allowed ranges for the
three mixing angles and the neutrino mass-squared differences are
given in Table~\ref{tab:nuparams}.  The phase $\delta$ and the mass of
the lightest neutrino are undetermined, although tritium beta decay
experiments set an upper limit on the neutrino masses of about
2~eV~\cite{Amsler:2008zzb}.

\begin{table}
\begin{tabular}{cc}
\hline \hline
Parameter & Value \\
\hline
$\sin^2 \theta_{12}$ & $0.314(1^{+0.18}_{-0.15})$ \\
$\sin^2 \theta_{23}$ & $0.44(1^{+0.41}_{-0.22})$ \\
$\sin^2 \theta_{13}$ & $0.9^{+2.3}_{-0.9} \times 10^{-2}$ \\
\hline
$\Delta m^2 \equiv m^2_{\nu_2} - m^2_{\nu_1}$ 
  & $7.92(1 \pm 0.09) \times 10^{-5}$~eV$^{2}$ \\
$\Delta M^2 \equiv m^2_{\nu_3} - \frac{1}{2}(m^2_{\nu_1} + m^2_{\nu_2})$ 
  & $\pm 2.4(1^{+0.21}_{-0.26}) \times 10^{-3}$~eV$^{2}$ \\
\hline \hline
\end{tabular}
\caption{Current values of the neutrino mixing parameters and
  mass-squared differences, from the global fit to neutrino
  oscillation data performed in Ref.~\cite{Fogli:2005cq}.
  Uncertainties quoted are the 2$\sigma$ ranges.  Note that the
  constraint on $\sin\theta_{13}$ is only an upper bound, and that the
  sign of $\Delta M^2$ is not yet known.}
\label{tab:nuparams}
\end{table}

Since the decays of $H^0$ and $A^0$ to two neutrinos will be invisible
to a collider detector, the decay of most interest is $H^+ \rightarrow
\ell^+ \nu$.  The charged Higgs can decay into all nine combinations
of $\ell_i \nu_j$; summing over neutrino mass eigenstates, the partial
width to a particular charged lepton $\ell$ is~\cite{Davidson:2009ha}
\begin{equation}
  \Gamma \left(H^+ \to \ell^+ \nu \right) 
  = \frac{M_{H^+} \langle m^2_{\nu} \rangle_{\ell}}{8\pi v_2^2},
\end{equation}
where we define the expectation value of the neutrino mass-squared in
a flavor eigenstate by~\cite{Fukuyama:2008sz}
\begin{equation}
  \langle m^2_{\nu} \rangle_{\ell} = \sum_i m^2_{\nu_{i}} |U_{\ell i}|^2.
\end{equation}

In what follows we work under the assumption that $M_{H^0, A^0} >
M_{H^+}$, i.e., $\lambda_4 > 0$, so that the decays $H^+ \rightarrow
W^+ H^0$, $W^+ A^0$ will be kinematically forbidden.  The branching
ratios of the charged Higgs are then completely determined by the
neutrino masses and mixing:
\begin{equation}
  {\rm BR}(H^+ \to \ell^+ \nu) 
  = \frac{\langle m^2_{\nu} \rangle_{\ell}}
  {\sum_{\ell} \langle m^2_{\nu} \rangle_{\ell}}
  = \frac{\langle m^2_{\nu} \rangle_{\ell}}
  {\sum_i m^2_{\nu_i}},
\end{equation}
where we used the unitarity of the PMNS matrix to simplify the
denominator.

\begin{figure}
\resizebox{1\textwidth}{!}{
{\rotatebox{270}{\includegraphics{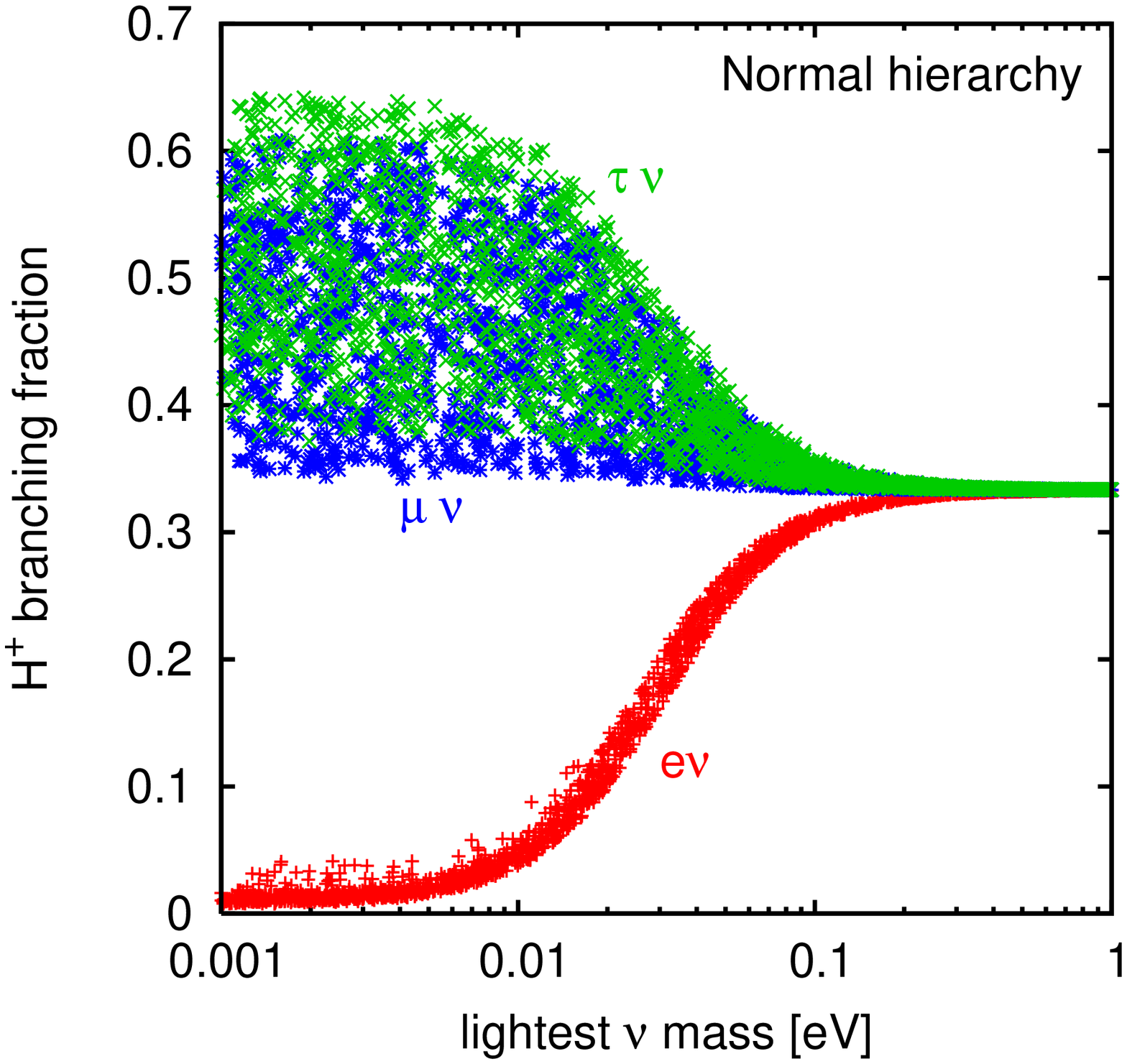}}}
{\rotatebox{270}{\includegraphics{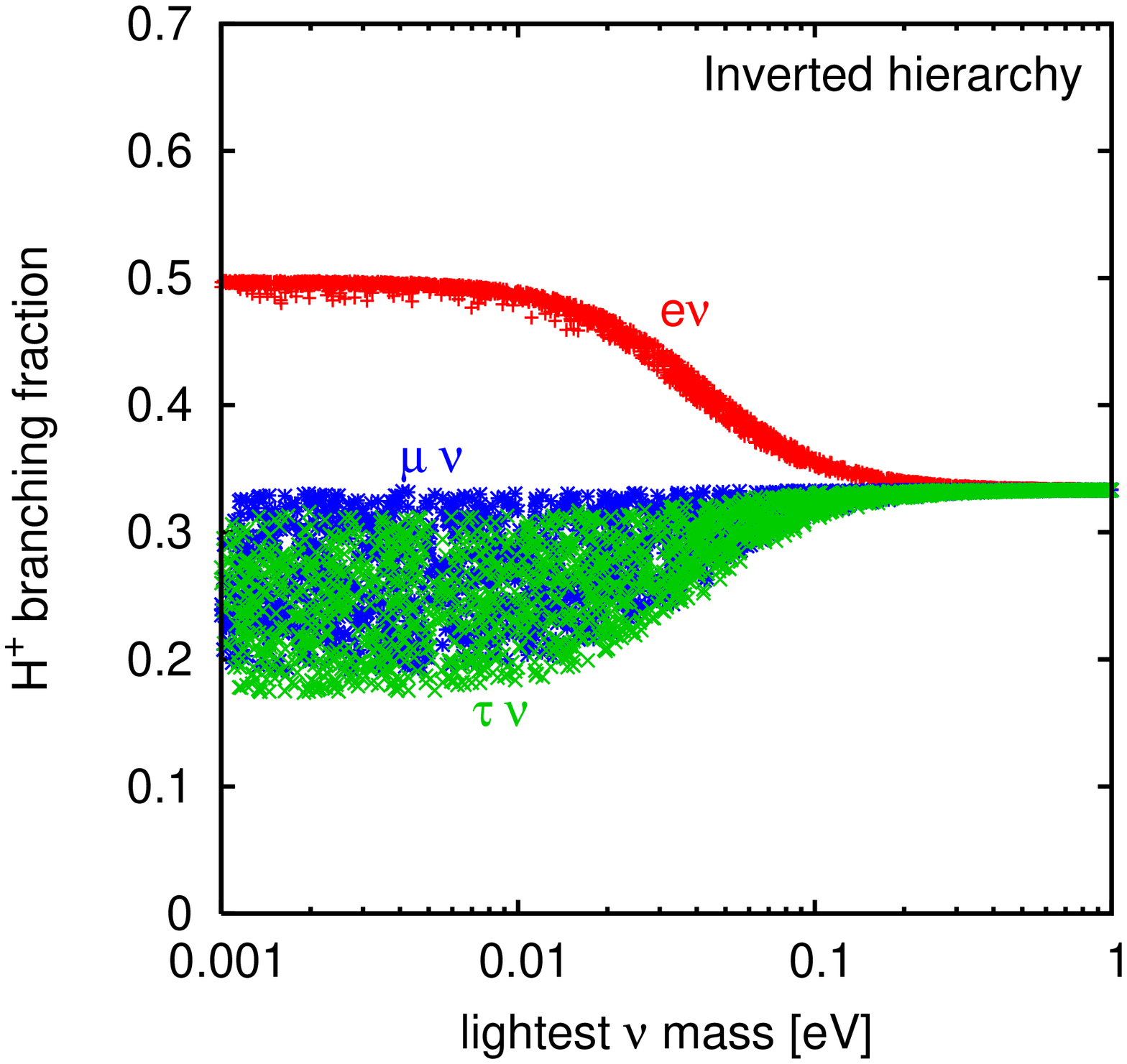}}}}
\caption{Charged Higgs decay branching fractions to $e\nu$, $\mu\nu$,
and $\tau\nu$ as a function of the lightest neutrino mass.}
\label{fig:brs}
\end{figure}

The sign of the larger neutrino mass splitting $\Delta M^2$ is unknown
(see Table~\ref{tab:nuparams}).  The situation in which $\Delta M^2$
is positive, so that $\nu_3$ is the heaviest neutrino, is called the
normal neutrino mass hierarchy, while the situation in which $\Delta
M^2$ is negative, so that $\nu_1$ and $\nu_2$ are heavier, is called
the inverted hierarchy.  We compute the charged Higgs branching
fractions as a function of the lightest neutrino mass for both
hierarchies, scanning over the 2$\sigma$ allowed ranges of the
neutrino parameters as given in Table~\ref{tab:nuparams}.  Results are
shown in Fig.~\ref{fig:brs}.\footnote{We disagree with the charged
  Higgs branching fractions to leptons presented in
  Ref.~\cite{Gabriel:2008es} for the $Z_2$ model of
  Ref.~\cite{Gabriel:2006ns}; these decays should have the same
  relative branching fractions as in our model.}  The large
spread in the branching ratios to $\mu \nu$ and $\tau \nu$ for
lightest neutrino masses below about 0.06~eV is due to the current
experimental uncertainty in $\sin^2\theta_{23}$, which controls the
relative amount of $\nu_{\mu}$ and $\nu_{\tau}$ in the isolated mass
eigenstate $\nu_3$.

Limits on the model parameters were discussed in
Ref.~\cite{Davidson:2009ha}.  The most significant for our purposes is
from searches for leptons plus missing energy at the CERN Large
Electron-Positron Collider, which put a lower bound on the charged
Higgs mass of 65--85 GeV, depending on the mass of the lightest
neutrino.  Big-bang nucleosynthesis also puts an upper bound on the
neutrino Yukawa couplings of
\begin{equation}
  y^{\nu}_i \equiv \frac{\sqrt{2} m_{\nu_i}}{v_2} \lesssim
  \frac{1}{30}\left[\frac{M_{H^+}}{100 \ {\rm GeV}}\right]
  \left[\frac{1/\sqrt{2}}{|U_{\ell i}|} \right].
\end{equation}

\section{Signal and background at the LHC}
\label{sec:SB}

In most other two-Higgs-doublet models, the charged Higgs decay rate
to a particular charged lepton is proportional to the square of the
charged lepton mass (see, e.g., Ref.~\cite{HHG}).  Such a charged
Higgs therefore decays predominantly to $\tau \nu$, with decays to
$\mu\nu$, $e \nu$ below 1\%.  In our neutrino-mass model, however, the
charged Higgs decay rate to a particular charged lepton is instead
proportional to the square of the mass of the corresponding neutrino
flavor eigenstate.  As a result, the branching fraction to $e \nu$
and/or $\mu \nu$ will always be sizable.  In particular, in the normal
hierarchy BR($H^+ \to \mu \nu) \simeq 1/2$, in the inverted hierarchy
BR($H^+ \to e \nu) \simeq 1/2$ and BR($H^+ \to \mu \nu) \simeq 1/4$,
and for a degenerate neutrino spectrum BR($H^+ \to e \nu) \simeq$
BR($H^+ \to \mu \nu) \simeq 1/3$, as shown in Fig.~\ref{fig:brs}.
Considering the high detection efficiency and lower fake rates of $e$
and $\mu$ compared to $\tau$, we study $H^+H^-$ pair production at the
LHC mediated by a photon or $Z$, followed by decays to $e$ or $\mu$
with missing transverse momentum.  We consider two scenarios, $M_{H^+}
= 100$ and 300~GeV, and present results as a function of the lightest
neutrino mass for both the normal and inverted hierarchy.

The process of interest is $pp \rightarrow H^+ H^- \rightarrow \ell
\ell^{\prime} \bar \nu_\ell \nu_{\ell^{\prime}}$, with $\ell
\ell^{\prime} = e^+ e^-$, $\mu^+ \mu^-$, and $e^{\pm} \mu^{\mp}$.  The
relevant backgrounds are $pp\rightarrow VV \rightarrow
\ell \ell^{\prime} \bar \nu \nu$ with $VV = W^+W^-$, $ZZ$, or $Z \gamma$
and the neutrinos of any type, and $pp \rightarrow t \bar t \rightarrow
\ell \ell^{\prime} \bar \nu_{\ell} \nu_{\ell^{\prime}} b \bar b$.  In
spite of the presence of the extra $b$ jets that can be vetoed, the $t
\bar t$ process is important because of its exceptionally high cross
section at the LHC.

\subsection{Event generation}
\label{sec:evtgen}

We simulated the signal and background processes with the parton-level
Monte Carlo MadGraph/MadEvent version~4~\cite{Alwall:2007st}.  We
present both a parton-level analysis and an analysis including
showering, hadronization, and a fast detector simulation using a
PYTHIA-PGS package designed to be used with MadEvent.  PYTHIA (version
6.4.20)~\cite{PYTHIA} generates initial- and final-state radiation and
hadronizes the final-state quarks and gluons, while PGS (Pretty Good
Simulation of High Energy Collisions, version 4)~\cite{PGS} is a basic
detector simulator---we used the default settings for ATLAS.  For the
signal process we generated 10,000 unweighted events in each of the
$e^+e^-$, $\mu^+ \mu^-$, and $\mu^+ e^-$ final states.  For both the
$VV$ and $t \bar t$ backgrounds we generated 100,000 unweighted events
in each of the three leptonic final states. We incorporated the $\mu^- e^+$ final state by
doubling the $\mu^+ e^-$ cross sections.  For the backgrounds we used
the default SM branching fractions from MadGraph/MadEvent, given in
Table~\ref{tab:WZbrs}.

\begin{table}
\begin{tabular}{cc}
\hline \hline
Process & Branching fraction \\
\hline
$W^+ \rightarrow \ell^+\nu_\ell$ ($\ell = e$ or $\mu$) & 0.1068 \\
$Z \rightarrow \ell^+ \ell^-$ ($\ell = e$ or $\mu$) & 0.0336 \\
$Z \rightarrow \nu \bar \nu$ (all 3 neutrinos) & 0.2000 \\
$t \rightarrow W^+b$ & 1.0000 \\
\hline \hline
\end{tabular}
\caption{Default SM branching fractions used in
  MadGraph/MadEvent~\cite{Alwall:2007st}.}
\label{tab:WZbrs}
\end{table}

Although MadGraph/MadEvent is a tree-level event generator, we
partially incorporated next-to-leading order (NLO) QCD corrections.
We did this for two reasons.  First, QCD corrections have a
significant effect on the signal and background (especially $t \bar
t$) cross sections, as well as significantly reducing the QCD scale
uncertainty, so that using NLO cross sections lets us obtain more
reliable results.  Second, for the $M_{H^+}=100$~GeV simulation we will apply a jet veto to reduce the $t
\bar t$ background, which will also affect the signal and $VV$
background once initial-state radiation is included.  While this could
be simulated by running the no-jet events through PYTHIA, a
parton-level simulation of the $H^+H^-j$ and $VVj$ processes provides
a more accurate description of jet kinematics.  Because these one-jet
processes make up part of the NLO QCD cross section for the
corresponding no-jet processes, we must incorporate the NLO cross
sections for consistency, as follows.

In the absence of a full NLO Monte Carlo, NLO QCD corrections are
usually incorporated by multiplying the leading-order (LO) cross
section---and the cross section corresponding to each simulated event
both before and after cuts---by a $k$-factor equal to the ratio of the
NLO cross section to the tree-level cross section.  In our case,
however, our jet veto will affect LO events (which have no jet) and
NLO events (which can have a final-state jet) differently.  We deal
with this by simulating $pp \to H^+H^- j$ and $pp \to VVj$ with the
same decay final states as considered in the no-jet processes.  For
simplicity we generate the same number of events with an additional
jet at the parton level as were generated for the no-jet processes.
Because the $t \bar t$ background already contains two jets at leading
order, we do not separately generate events with additional jets for
this background.  To avoid the collinear and infrared singularities,
we apply a minimum $p_T$ cut of 10~GeV on the jet at the
event-generation level.

The square of the NLO matrix element can be expressed up to order
$\alpha_s$ as
\begin{equation}
  \mathcal{|M|}_{\rm NLO}^2 
  = | \mathcal{M}_{\rm LO} + \mathcal{M}_{\rm 1 \, loop} |^2 
  + \mathcal{|M|}_{\rm 1 \, jet}^2.
\end{equation}
We used MadGraph/MadEvent to calculate the cross sections
corresponding to $\mathcal{M}_{\rm LO}$ and $\mathcal{M}_{\rm 1 \,
  jet}$.  We computed the NLO cross-section for $pp \to H^+H^-$ at the
LHC using the public FORTRAN code {\tt
  PROSPINO}~\cite{Beenakker:1999xh,Alves:2005kr} with CTEQ6 parton
densities~\cite{Pumplin:2002vw}, with the renormalization and
factorization scales set equal to $M_{H^+}$.  We took the NLO cross
sections for the SM $W^+W^-$ and $ZZ$ background processes from
Ref.~\cite{Campbell:1999ah}.  This paper quotes results using both the
MRS98 and CTEQ5 parton densities, with results differing by $\sim$6\%;
since we use CTEQ6 for the tree-level MadGraph/MadEvent calculation,
we take the results using the CTEQ5 parton densities for consistency.
For events with $e^{\pm} \mu^{\mp}$ in the final state, only the cross
section for $W^+W^-$ is relevant; for events with $\mu^+\mu^-$ or
$e^+e^-$ in the final state, both the $W^+W^-$ and $ZZ$ processes
contribute and we add the cross sections at both LO and NLO.  We took
the $t \bar t$ cross section from Ref.~\cite{Bonciani:1998vc}, which
includes both NLO and next-to-leading logarithmic corrections.
The remaining scale uncertainty is about $\pm$5\% when the
factorization and renormalization scales are varied between $m_t/2$
and $2m_t$.  The relevant cross sections are given in
Table~\ref{tab:xsecs}.  

\begin{table}
\begin{tabular}{ccc}
\hline \hline
Process & Cross section & Source \\
\hline
$pp \to H^+H^-$ ($M_{H^+} = 100$~GeV) & 295 fb 
   & {\tt PROSPINO}~\cite{Beenakker:1999xh,Alves:2005kr} \\
$pp \to H^+H^-$ ($M_{H^+} = 300$~GeV) & 5.32 fb
   & {\tt PROSPINO}~\cite{Beenakker:1999xh,Alves:2005kr} \\
$pp \to W^+W^-$ & 127.8 pb & Ref.~\cite{Campbell:1999ah} \\
$pp \to ZZ$ & 17.2 pb & Ref.~\cite{Campbell:1999ah} \\
$pp \to t \bar t$ & 833 pb & Ref.~\cite{Bonciani:1998vc} \\
\hline \hline
\end{tabular}
\caption{NLO cross sections for signal and background processes (before
  decays) at the LHC (14~TeV).  The $t \bar t$ cross section also
  includes a resummation of next-to-leading logarithmic corrections.}
\label{tab:xsecs}
\end{table}

We find that with our generator-level jet
$p_T$ cut on $\sigma_{\rm 1 \, jet}$, $\sigma_{\rm NLO} < \sigma_{\rm
  LO} + \sigma_{\rm 1 \, jet}$, so the one-loop matrix element must
interfere destructively with the LO matrix element.  Thus the
generated cross section from the LO process must be scaled down in
order to incorporate the effects of the one-loop correction.
For the parton-level simulation, the relevant scale factor is
determined by solving for $k$ in the equation,
\begin{equation}
  \sigma_{\rm NLO} = k \, \sigma_{\rm LO} + \sigma_{\rm 1 \, jet},
  \label{eq:k}
\end{equation}
before cuts are applied, and then using this equation with the same
value of $k$ to calculate the surviving $\sigma_{\rm NLO}$ after the
cuts are applied to the LO and one-jet MadGraph/MadEvent simulated
results.

For the PYTHIA-PGS simulation, the simulated events have extra jets
produced by PYTHIA and ``measured'' jet $p_T$ smeared by PGS.  To
avoid double-counting, we use the following equation with two
constants:
\begin{equation}
  \sigma_{\rm NLO} = m \, \sigma_{\rm LO}^{\rm cut} 
  + n \, \sigma_{\rm 1 \, jet}^{\rm cut},
  \label{eq:mn}
\end{equation}
where $\sigma_{\rm LO}^{\rm cut}$ and $\sigma_{\rm 1 \, jet}^{\rm
  cut}$ are the cross sections identified by PGS as having no jets and
at least one jet, respectively, with $p_T > 10$ GeV, out of the
combined LO and one-jet generated samples.  The constants $m$ and $n$
are determined by $m \, \sigma_{\rm LO}^{\rm cut} = k \, \sigma_{\rm
  LO}$ and $n \, \sigma_{\rm 1 \, jet}^{\rm cut} = \sigma_{\rm 1 \,
  jet}$ using $k$ from Eq.~(\ref{eq:k}).  Equation~(\ref{eq:mn}) with
the same values of $m$ and $n$ is then used after cuts to calculate
the surviving $\sigma_{\rm NLO}$.

\subsection{Cuts}

We apply four cuts to reduce the background, summarized in
Table~\ref{tab:cuts}.  The first cut checks for the presence of two
opposite-sign leptons each with $p_T > 20$~GeV and missing transverse
momentum of at least 30~GeV.  For the parton-level simulation, we also
apply acceptance cuts on the pseudorapidity of both leptons, $|\eta| <
3.0$ for electrons and $|\eta| < 2.4$ for muons.  Second, for the
$e^+e^-$ and $\mu^+\mu^-$ final states we veto events for which the
dilepton invariant mass falls between 80 and 100~GeV, in order to
eliminate background from $Z (\to \ell \ell) + p_T^{\rm miss}$.  This
will also eliminate the majority of any background from $Z + jets$
with fake $p_T^{\rm miss}$, which we did not simulate.  The third cut
vetoes events containing a jet with $p_T > 30$~GeV; for the
parton-level simulation, we require that this jet falls in the
rapidity range $|\eta| < 5.0$.  This eliminates more than 97\% of the
$t \bar t$ backgound, but also reduces the signal by about a factor of
two.  We find that this cut is useful for $M_{H^+} = 100$~GeV.  For
$M_{H^+} = 300$~GeV the signal cross section is considerably smaller
and the signal events will be better separated from background in our
final cut variable, so that we obtain better sensitivity without the
jet veto.

\begin{table}
\begin{tabular}{lp{13cm}}
\hline \hline 
Cut name & Explanation \\ 
\hline 
Basic cuts & Present are a lepton and antilepton, each with
$p_T^{\ell} > 20$~GeV, and missing transverse momentum $p_T^{\rm miss}
> 30$~GeV.  For the parton level results, we also apply lepton
acceptance cuts of $|\eta_e|<3.0$ and $|\eta_\mu|<2.4$.\\
$Z$ pole veto \ \ & To eliminate events that include $Z \rightarrow
\ell^+\ell^-$, we veto events in which the invariant mass of $e^+e^-$
or $\mu^+\mu^-$ is between 80 and 100~GeV (not applied to the
$e^{\pm}\mu^{\mp} p_T^{\rm miss}$ final state).\\
Jet veto & Designed to reduce $t \bar t$ background, any event with a
jet with $p_T^{\rm jet} > 30$~GeV was rejected.  For the parton level
results, this veto is only applied when $|\eta_{\rm jet}| < 5.0$.
(Applied only for $M_{H^+} = 100$~GeV.)\\
$M_{T2}$ cut & To reduce the $W^+W^-$ and $t \bar t$ backgrounds, we
make use of the larger mass of $H^+$ compared to the intermediate $W$
bosons in both backgrounds by cutting on $M_{T2}$ (defined in
Eq.~(\ref{eq:mT2})).  For $M_{H^+} = 100$~GeV we require $M_W < M_{T2}
< 100$~GeV and for $M_{H^+} = 300$~GeV we require 150~GeV~$< M_{T2} <
300$~GeV.\\
\hline \hline
\end{tabular}
\caption{Summary of cuts.}
\label{tab:cuts}
\end{table}

The final cut is on the variable $M_{T2}$, defined as~\cite{Lester:1999tx}
\begin{equation}
  M_{T2}^2 = \begin{array}{c} \rm min \\ 
    q_T^{\rm miss(1)} + q_T^{\rm miss(2)} = p_T^{\rm miss} \end{array} 
  \left[ {\rm max} \left\{ m_T^2 \left( p_T^{\ell(1)}, q_T^{\rm miss(1)} 
    \right) , 
    m_T^2 \left(p_T^{\ell(2)}, q_T^{\rm miss(2)} \right) \right\} \right],
\label{eq:mT2}
\end{equation}
where $m_T^2$ is the square of the transverse mass (ignoring the charged
lepton and neutrino masses),
\begin{equation}
  m_T^2 \left( p_T^{\ell}, q_T^{\rm miss} \right) 
  = 2 \left( |\vec p_T^{\, \ell}| |\vec q_T^{\, \rm miss}| 
  - \vec p_T^{\, \ell} \cdot \vec q_T^{\, \rm miss} \right).
\end{equation}
In other words, $M_{T2}$ is determined by making a guess for the
transverse momenta of the two neutrinos (constrained by the measured
total missing transverse momentum) and computing the transverse masses
of the two $\ell \nu$ systems; the guess is then varied until the
larger of the two reconstructed transverse masses is minimized. 

For equal-mass intermediate particles each decaying to $\ell \nu$, the
$M_{T2}$ distribution has an upper endpoint at the mass of the
intermediate particle.  Thus by cutting out events with $M_{T2} <
M_W$, all the $W^+W^-$ background should be eliminated (the endpoint
is in fact smeared out by the finite $W$ width and momentum resolution
of the detector).  Since the leptons and missing transverse momentum
in the $t \bar t$ process also come from decays of on-shell $W^+W^-$,
this background should be eliminated as well.  There is also a
small contribution to the $VV$ background from nonresonant processes
that can have $M_{T2} > M_W$.  Since all signal events will have
$M_{T2} < M_{H^+}$, we also cut out events with $M_{T2} > M_{H^+}$ in
an effort to reduce the background from these nonresonant $VV$
processes.  For $M_{H^+}=300$~GeV, we find that raising the minimum
cut on $M_{T2}$ to 150~GeV reduces the tail of the nonresonant $VV$
events without reducing the signal too much.  

\begin{figure}
\resizebox{\textwidth}{!}{
\includegraphics{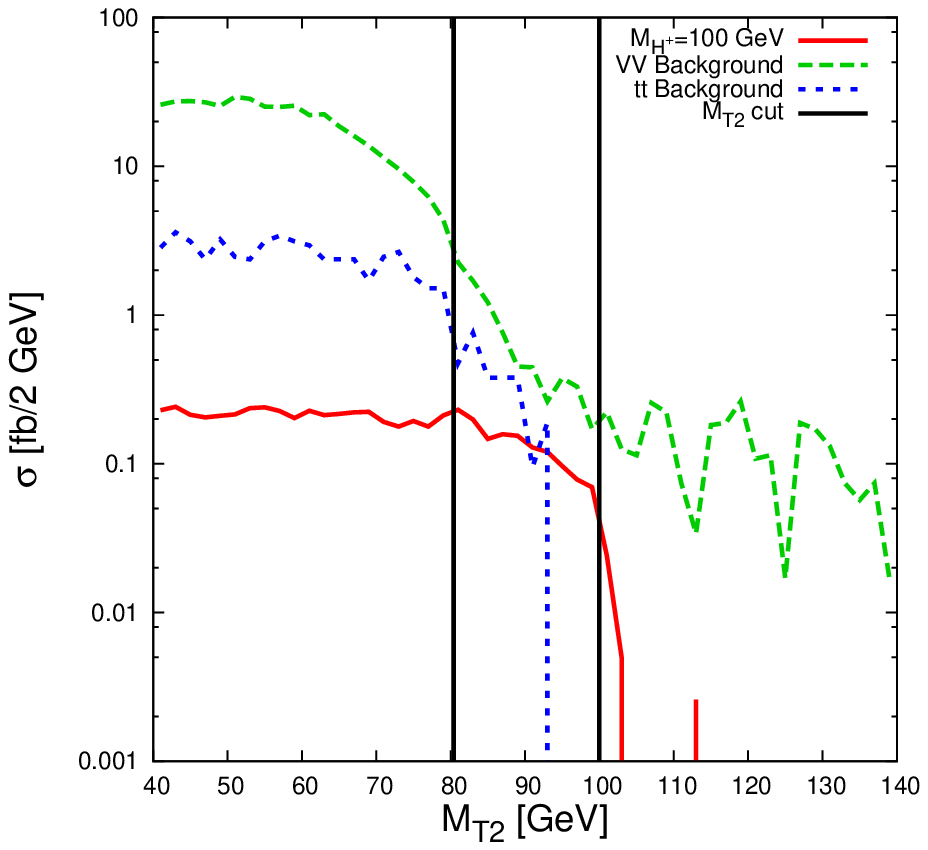}
\includegraphics{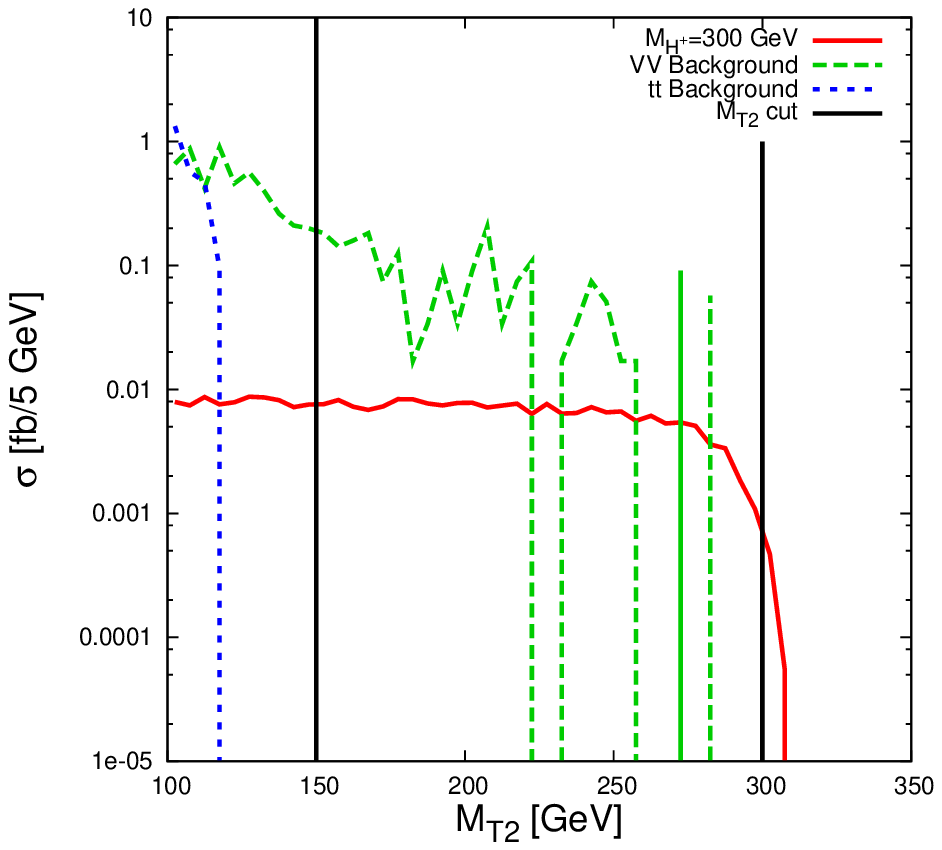}}
\caption{$M_{T2}$ distributions after other cuts have been applied for
  the $e^+ e^- p_T^{\rm miss}$ final state, with $M_{H^+}=100$~GeV
  (left, with jet veto) and 300~GeV (right, no jet veto).  For the
  signal we take BR($H^+ \to e^+ \nu) = 1/3$, which occurs for a
  degenerate neutrino spectrum.  The $M_{T2}$ cut window is shown by
  the vertical lines.}
\label{fig:hist}
\end{figure}

In Fig.~\ref{fig:hist} we show the $M_{T2}$ distributions for signal
and background processes in the $e^+e^- p_T^{\rm miss}$ channel for
$M_{H^+}=100$ and 300~GeV after the other cuts have been applied, for
the PYTHIA-PGS simulation.  Note that the $t \bar t$ background
distribution has a maximum $M_{T2}$ value a little above the $W$ mass,
so that it can be eliminated with a high enough cut on $M_{T2}$, as we
do for the case of $M_{H^+} = 300$~GeV.  (The higher $M_{T2}$ endpoint
for $t \bar t$ in the right-hand plot in Fig.~\ref{fig:hist} is due to
the absence of the jet veto, resulting in much higher $t \bar t$
statistics.)  The $VV$ background also falls off dramatically around
$M_{T2} \sim M_W$; however, due to nonresonant diagrams without
on-shell intermediate $W$ pairs, this background extends to much
higher values of $M_{T2}$.  With our simulation statistics, a single
$\bar t t$ event corresponds to a cross section of about 0.1~fb, while
a single $VV$ event corresponds to a cross section of about 0.01~fb.

\subsection{Results}

The efficiency of each cut on $\sigma_{\rm NLO}$ for the $e^+ e^-
p_T^{\rm miss}$ final state is displayed in Tables~\ref{eehh},
\ref{eebg100} and~\ref{eebg300}. Cut efficiencies for $\mu^+
\mu^-p_T^{\rm miss}$ are displayed in Tables~\ref{mmhh}, \ref{mmbg100}
and~\ref{mmbg300}, and for $e^\pm \mu^\mp p_T^{\rm miss}$ in
Tables~\ref{emhh}, \ref{embg100} and~\ref{embg300}.  We give
efficiencies for both the parton-level simulation and the simulation
including showering, hadronization, and fast detector simulation using
the PYTHIA-PGS package.  All results incorporate NLO corrections as
described in Sec.~\ref{sec:evtgen}.

\begin{table} 
\begin{tabular}{lcccc} 
\hline \hline
&\multicolumn{2}{c}{$M_{H^+}=100$~GeV}&\multicolumn{2}{c}{$M_{H^+}=300$~GeV}\\
Cuts&Parton&PYTHIA/PGS&Parton&PYTHIA/PGS \\ 
\hline 
Basic cuts&0.62644&0.47860&0.92961&0.72105\\ 
$Z$ pole veto&0.90754&0.90165&0.97732&0.97724\\ 
Jet veto&0.68433&0.60717&--&--\\ 
$M_{T2}$ cut&0.17075&0.15542&0.47317&0.45945\\ 
\hline
Cumulative&0.06643&0.04072&0.42988&0.32375\\ 
\hline \hline
\end{tabular}
\caption{Cut efficiencies for the signal process $pp \rightarrow e^+
e^- p_T^{\rm miss}$ via $H^+ H^-$.  The efficiency of each cut is
defined as the cross section that passed the cut divided by the
cross section that passed the previous cut.  The cumulative efficiency
is the cross section that passed all the cuts divided by the original
cross section.  The jet veto is not applied for $M_{H^+} = 300$~GeV.
The $M_{T2}$ cut is $M_W < M_{T2} < 100$~GeV for
$M_{H^+}=100$~GeV, and 150~GeV $ < M_{T2} < 300$~GeV for $M_{H^+}=300$~GeV.}  
\label{eehh} 
\end{table}

\begin{table}
\begin{tabular}{lcccc}
\hline \hline
&\multicolumn{2}{c}{$VV$ Background}&\multicolumn{2}{c}{$t \bar t$ Background}\\
Cuts&Parton&PYTHIA/PGS&Parton&PYTHIA/PGS \\
\hline
Basic cuts&0.42708&0.33912&0.58407&0.40612\\
$Z$ pole veto&0.74727&0.73255&0.86236&0.85501\\
Jet veto&0.63306&0.67299&0.01318&0.02856\\
$M_{W^+} < M_{T2} < 100$ GeV&0.01401&0.01147&0.01205&0.02716\\
\hline
Cumulative&0.00283&0.00192&0.00008&0.00027\\
\hline \hline
\end{tabular}
\caption{As in Table~\ref{eehh} but for background for 
$pp \rightarrow e^+ e^- p_T^{\rm miss}$, with cuts for $M_{H^+}=100$~GeV.}
\label{eebg100}
\end{table}

\begin{table}
\begin{tabular}{lcccc}
\hline \hline
&\multicolumn{2}{c}{$VV$ Background}&\multicolumn{2}{c}{$t \bar t$ Background}\\
Cuts&Parton&PYTHIA/PGS&Parton&PYTHIA/PGS \\
\hline
Basic cuts&0.42708&0.33912&0.58407&0.40612\\
$Z$ pole veto&0.74727&0.73255&0.86236&0.85501\\
150 GeV $ < M_{T2} < 300$ GeV&0.00260&0.00196&0.00000&0.00000\\
\hline
Cumulative&0.00083&0.00049&0.00000&0.00000\\
\hline \hline
\end{tabular}
\caption{As in Table~\ref{eehh} but for background for 
$pp \rightarrow e^+ e^- p_T^{\rm miss}$, with cuts for $M_{H^+}=300$~GeV.}
\label{eebg300}
\end{table}

\begin{table}
\begin{tabular}{lcccc}
\hline \hline
&\multicolumn{2}{c}{$M_{H^+}=100$ GeV}&\multicolumn{2}{c}{$M_{H^+}=300$ GeV}\\
Cuts&Parton&PYTHIA/PGS&Parton&PYTHIA/PGS \\
\hline
Basic cuts&0.51713&0.43680&0.84810&0.69845\\
$Z$ pole veto&0.90869&0.90075&0.97756&0.97696\\
Jet veto&0.68310&0.57831&--&--\\
$M_{T2}$ cut&0.16875&0.17320&0.47802&0.46847\\
\hline
Cumulative&0.05417&0.03941&0.39632&0.31966\\
\hline \hline
\end{tabular}
\caption{As in Table~\ref{eehh} but for the signal process
$pp \rightarrow \mu^+ \mu^- p_T^{\rm miss}$ via $H^+ H^-$.}
\label{mmhh}
\end{table}

\begin{table}
\begin{tabular}{lcccc}
\hline \hline
&\multicolumn{2}{c}{$VV$ Background}&\multicolumn{2}{c}{$t \bar t$ Background}\\
Cuts&Parton&PYTHIA/PGS&Parton&PYTHIA/PGS \\
\hline
Basic cuts&0.32959&0.28226&0.52593&0.39048\\
$Z$ pole veto&0.73839&0.73098&0.86021&0.85489\\
Jet veto&0.62703&0.63204&0.01346&0.02282\\
$M_{W^+} < M_{T2} < 100$ GeV&0.01324&0.01554&0.00657&0.03675\\
\hline
Cumulative&0.00202&0.00203&0.00004&0.00028\\
\hline \hline
\end{tabular}
\caption{As in Table~\ref{eehh} but for background for 
$pp \rightarrow \mu^+ \mu^- p_T^{\rm miss}$, with cuts for $M_{H^+}=100$~GeV.}
\label{mmbg100}
\end{table}

\begin{table}
\begin{tabular}{lcccc}
\hline \hline
&\multicolumn{2}{c}{$VV$ Background}&\multicolumn{2}{c}{$t \bar t$ Background}\\
Cuts&Parton&PYTHIA/PGS&Parton&PYTHIA/PGS \\
\hline
Basic cuts&0.32959&0.28226&0.52593&0.39048\\
$Z$ pole veto&0.73839&0.73098&0.86021&0.85489\\
150 GeV $ < M_{T2} < 300$ GeV&0.00288&0.00239&0.00000&0.00000\\
\hline
Cumulative&0.00070&0.00049&0.00000&0.00000\\
\hline \hline
\end{tabular}
\caption{As in Table~\ref{eehh} but for background for 
$pp \rightarrow \mu^+ \mu^- p_T^{\rm miss}$, with cuts for $M_{H^+}=300$~GeV.}
\label{mmbg300}
\end{table}

\begin{table}
\begin{tabular}{lcccc}
\hline \hline
&\multicolumn{2}{c}{$M_{H^+}=100$ GeV}&\multicolumn{2}{c}{$M_{H^+}=300$ GeV}\\
Cuts&Parton&PYTHIA/PGS&Parton&PYTHIA/PGS \\
\hline
Basic cuts&0.56131&0.45743&0.88249&0.70832\\
Jet veto&0.68783&0.59528&--&--\\
$M_{T2}$ cut&0.16857&0.16121&0.47427&0.46373\\
\hline
Cumulative&0.06508&0.04390&0.41854&0.32847\\
\hline \hline
\end{tabular}
\caption{As in Table~\ref{eehh} but for the signal process 
$pp \rightarrow e^\pm \mu^\mp p_T^{\rm miss}$ via $H^+ H^-$.}
\label{emhh}
\end{table}

\begin{table}
\begin{tabular}{lcccc}
\hline \hline
&\multicolumn{2}{c}{$VV$ Background}&\multicolumn{2}{c}{$t \bar t$ Background}\\
Cuts&Parton&PYTHIA/PGS&Parton&PYTHIA/PGS \\
\hline
Basic cuts&0.35835&0.30423&0.55297&0.39556\\
Jet veto&0.65590&0.68572&0.01255&0.02592\\
$M_{W^+} < M_{T2} < 100$ GeV&0.00860&0.01207&0.01585&0.03018\\
\hline
Cumulative&0.00202&0.00252&0.00011&0.00031\\
\hline \hline
\end{tabular}
\caption{As in Table~\ref{eehh} but for background for 
$pp \rightarrow e^\pm \mu^\mp p_T^{\rm miss}$, with cuts for 
$M_{H^+}=100$~GeV.}
\label{embg100}
\end{table}

\begin{table}
\begin{tabular}{lcccc}
\hline \hline
&\multicolumn{2}{c}{$VV$ Background}&\multicolumn{2}{c}{$t \bar t$ Background}\\
Cuts&Parton&PYTHIA/PGS&Parton&PYTHIA/PGS \\
\hline
Basic cuts&0.35835&0.30423&0.55297&0.39556\\
150 GeV $ < M_{T2} < 300$ GeV&0.00057&0.00049&0.00000&0.00000\\
\hline
Cumulative&0.00021&0.00015&0.00000&0.00000\\
\hline \hline
\end{tabular}
\caption{As in Table~\ref{eehh} but for background for 
$pp \rightarrow e^\pm \mu^\mp p_T^{\rm miss}$, with cuts for 
$M_{H^+}=300$~GeV.}
\label{embg300}
\end{table}

Consider for example the PYTHIA-PGS results in the $e^+e^- p_T^{\rm
  miss}$ final state, and assume a degenerate neutrino spectrum so
that BR($H^+ \to e^+ \nu) = 1/3$.  In this case, for $M_{H^+} =
100$~GeV, the cuts reduce the charged Higgs signal cross section in
this channel from 32.8~fb to 1.34~fb, while reducing the $VV$
background from 1570~fb to 3.01~fb and the $t \bar t$ background from
9500~fb to 2.57~fb.  The ratio of signal to background cross sections
(S/B) is then 0.24.  For $M_{H^+} = 300$~GeV, S/B is comparable.
These are displayed for all channels for a degenerate neutrino
spectrum in Table~\ref{SB}.  In all cases S/B is at least 0.22,
comfortably larger than the QCD and parton density uncertainties on
the $VV$ and $t \bar t$ backgrounds; the overall cross sections of
these backgrounds can also be normalized experimentally using $M_{T2}$
regions below $M_W$.

\begin{table}
\begin{tabular}{cccc}
\hline \hline
$M_{H^+}$&Channel& S/B & \ Luminosity for 5$\sigma$ \ \\
\hline
& \ $e^+e^-p_T^{\rm miss}$ \ & 0.24 & 78 fb$^{-1}$ \\
100~GeV&$\mu^+\mu^-p_T^{\rm miss}$& 0.22 & 88 fb$^{-1}$ \\
&$e^\pm \mu^\mp p_T^{\rm miss}$& 0.22 & 40 fb$^{-1}$ \\
\hline
&$e^+e^-p_T^{\rm miss}$& 0.25 & 526 fb$^{-1}$ \\
300~GeV&$\mu^+\mu^-p_T^{\rm miss}$& 0.25 & 540 fb$^{-1}$ \\
&$e^\pm \mu^\mp p_T^{\rm miss}$& 0.89 & 73 fb$^{-1}$ \\
\hline \hline
\end{tabular}
\caption{Signal over background (S/B) and luminosity required for a
  5$\sigma$ discovery in a single channel for the three signal
  processes studied, for $M_{H^+} = 100$ and 300~GeV, assuming a
  degenerate neutrino spectrum so that BR($H^+ \to e^+ \nu) = {\rm
    BR}(H^+ \to \mu^+ \nu) = 1/3$.}
\label{SB}
\end{table}

For $M_{H^+} = 100$~GeV, the background after cuts depends sensitively
on the shape of the background $M_{T2}$ distribution just above $M_W$.
This is controlled by the $W$ width and the detector resolution for
lepton momenta and missing $p_T$; its shape should not suffer from QCD
or parton-density uncertainties.  For $M_{H^+} = 300$~GeV, the shape
and normalization of the nonresonant tail of the $VV$ background is
especially important.  This background is mostly Drell-Yan with an
additional on-shell $W$ boson radiated from one of the final-state
leptons; the QCD corrections to such processes are well understood.
Given enough statistics, the shape of this background could also be
normalized using the $M_{T2}$ region above $M_{H^+}$.  Note also that
the nonresonant tail of the $VV$ background is significantly smaller
for the $e^{\pm} \mu^{\mp}$ final state than for the $e^+e^-$ and
$\mu^+\mu^-$ final states, leading to a much higher signal purity in
this final state for $M_{H^+}=300$~GeV as shown in the last line of
Table~\ref{SB} (for the lower charged Higgs mass this effect is
swamped by the resonant-$W$ contribution).

\begin{figure}
\resizebox{\textwidth}{!}{
\includegraphics{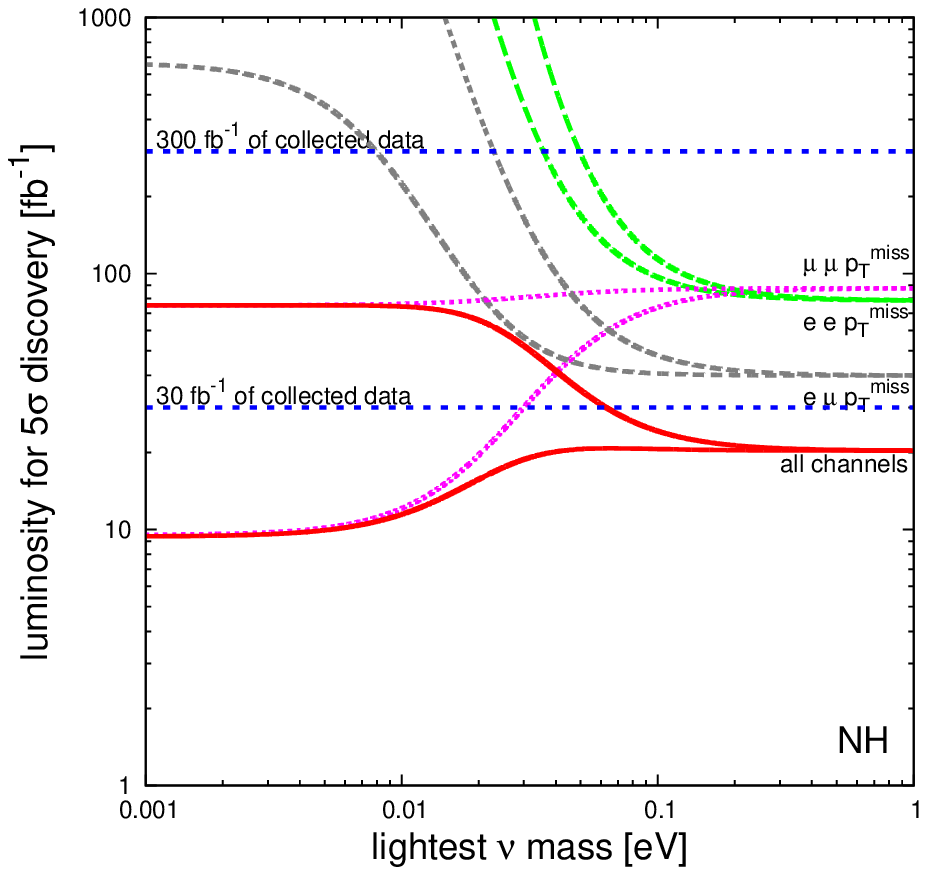}
\includegraphics{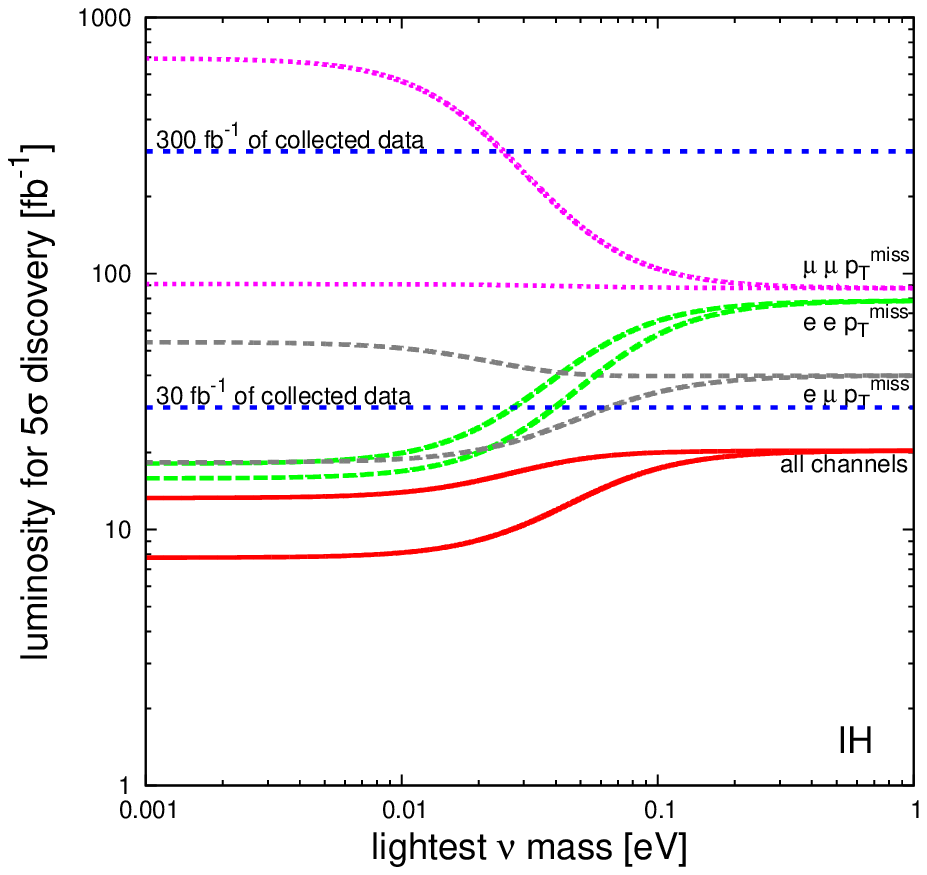}}
\caption{Luminosity required at the LHC (14~TeV) for a 5$\sigma$
  discovery if $M_{H^+} = 100$~GeV, for the normal hierarchy (NH,
  left) and inverted hierarchy (IH, right).  The lines for each
  channel bound the range of required luminosities obtained by
  scanning over the 2$\sigma$ allowed ranges of the parameters of the
  neutrino mixing matrix and mass-squared differences.}
\label{sig100}
\end{figure}

\begin{figure}
\resizebox{\textwidth}{!}{
\includegraphics{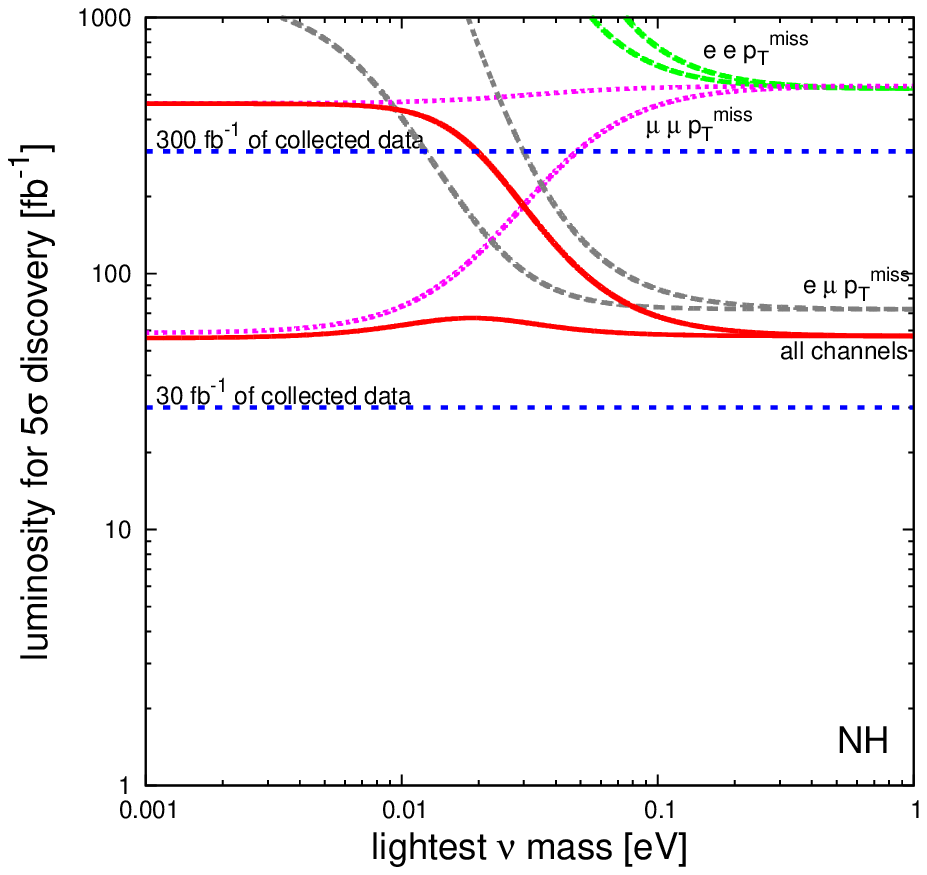}
\includegraphics{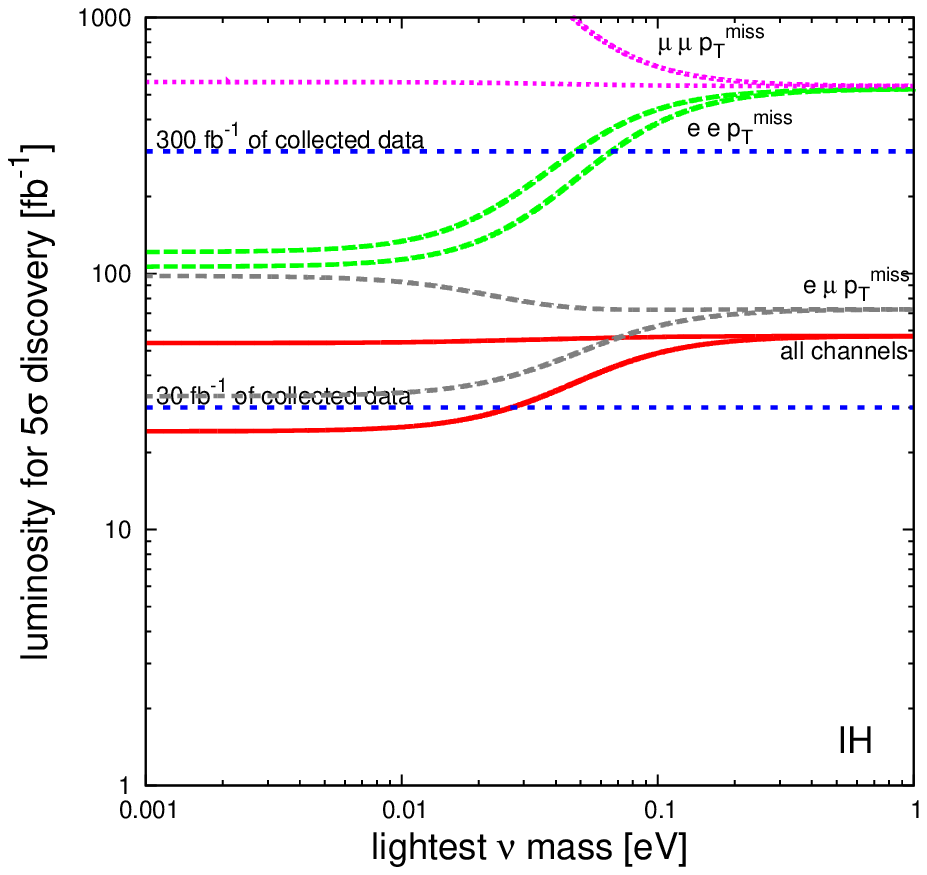}}
\caption{As in Fig.~\ref{sig100} but for $M_{H^+} = 300$~GeV.}
\label{sig300}
\end{figure}

The integrated luminosity required for a 5$\sigma$ discovery of
$H^+H^-$ is displayed in Fig.~\ref{sig100} for $M_{H^+} = 100$~GeV and
Fig.~\ref{sig300} for $M_{H^+} = 300$~GeV, for each channel separately
and for all three channels combined.  We use the PYTHIA-PGS results
and compute only the statistical significance.  For the normal
hierarchy with $M_{H^+} = 100$ (300)~GeV, we find 5$\sigma$ discovery
statistics with a minimum of 9 (56)~fb$^{-1}$.  For the inverted
hierarchy, the minimum is 8 (24)~fb$^{-1}$.  For the case of
degenerate neutrino masses, 20 (57)~fb$^{-1}$ is needed.  For
degenerate neutrino masses, the luminosity needed for a 5$\sigma$
discovery in each channel separately is given in Table~\ref{SB}.

\section{Discussion and conclusions}
\label{sec:conclusions}

The two-Higgs-doublet model for Dirac neutrino masses studied here
provides distinctive leptonic signatures at the LHC due to the
characteristic decay pattern of the charged Higgs boson, controlled by
the neutrino masses and mixing.  We have shown that a simple set of
cuts allows discovery of charged Higgs pairs with decays to
$\ell \ell^{(\prime)} p_T^{\rm miss}$ with relatively modest
integrated luminosity.  In particular we found that a cut on the
kinematic variable $M_{T2}$ provides very effective suppression of $W$
pair and $t \bar t$ backgrounds for charged Higgs masses sufficiently
above the $W$ mass.

In the inverted neutrino mass hierarchy, the large branching fractions
of the charged Higgs to $e \nu$ and $\mu \nu$ guarantees a 5$\sigma$
discovery for any allowed neutrino mass and mixing parameter values
with only 20 (57) fb$^{-1}$ for $M_{H^+} = 100$ (300)~GeV.  The
discovery potential remains remarkably good at $M_{H^+} = 300$~GeV
despite the rapidly falling charged Higgs pair production cross
section because of the increasing separation of the signal $M_{T2}$
distribution from the background.

In the normal neutrino mass hierarchy, the large uncertainty on the
neutrino mixing angle $\theta_{23}$ leads to parameter regions in
which the charged Higgs decays predominantly to $\tau\nu$, with a
branching fraction to light leptons below 40\%, resulting in poor
discovery sensitivity in the light lepton channels studied in this
paper.  Away from these parameter regions, the discovery prospects are
only slightly worse than in the inverted hierarchy.

As more stringent experimental limits are placed on the neutrino
parameters from neutrino oscillation experiments and direct searches
for the kinematic neutrino mass in beta decay, the predictions for the
charged Higgs branching ratios in this model will tighten.  For
example, one goal of the currently-running T2K long-baseline neutrino
oscillation experiment in Japan is to improve the measurement accuracy
of $\sin^2(2 \theta_{23})$ by an order of
magnitude~\cite{Zito:2008zza}, which would reduce the 2$\sigma$ spread
in the charged Higgs branching ratios to $\mu\nu$ and $\tau\nu$ at low
lightest-neutrino mass from the current $\pm$30\% to about $\pm$10\%.
Sensitivity to the neutrino mass hierarchy relies on detection of a
nonzero $\theta_{13}$, a major goal of T2K and the longer-baseline
U.S.-based experiment NO$\nu$A currently under
construction~\cite{NOvATDR}.  The ratios of the signal rates in the
three channels considered here would allow the normal, inverted, and
degenerate neutrino spectra to be differentiated, providing a key test
of the connection of the model to the neutrino sector.

Measurement of the charged Higgs branching fractions will also provide
some sensitivity to the mass of the lightest neutrino.  For a lightest
neutrino mass between about 0.01 and 0.1~eV, the charged Higgs
branching ratios vary dramatically with the value of the lightest
neutrino mass (Fig.~\ref{fig:brs}); once the measurement of
$\theta_{23}$ from neutrino oscillations has improved, measurement of
the ratio of the $e \nu$ and $\mu\nu$ modes will provide sensitivity
to the lightest neutrino mass in this range.  This is nicely
complementary to the prospects for direct kinematic neutrino mass
determination from the tritium beta decay experiment KATRIN, which is
designed to be sensitive down to neutrino masses of about
0.2~eV~\cite{Valerius:2005aw}---i.e., at the lower end of the
degenerate part of the spectrum---and is scheduled to begin
commissioning in 2012~\cite{KatrinTalk}.  We note that, because the
neutrinos in this model are Dirac particles, neutrinoless double beta
decay experiments will have no signal and will thus not be sensitive
to the neutrino mass scale.

The mass of the charged Higgs is also accessible at the LHC through
the signal event kinematics.  In particular, the signal $M_{T2}$
distribution is flat up to an endpoint at the charged Higgs mass, as
shown in Fig.~\ref{fig:hist}.  A fit to this distribution on top of
the background should provide a measurement of the charged Higgs mass.
This would allow a valuable cross-check of the charged Higgs pair
production cross section together with the visible branching fractions
as predicted by the neutrino parameters.  The pair production cross
section is sensitive to the isospin of the charged Higgs through its
coupling to the $Z$ boson, allowing the two-doublet nature of the
model to be established~\cite{Davidson:2009ha}.

We finally comment on the applicability of our results to two other
neutrino mass models that contain a charged Higgs boson.  First, the
$Z_2$ model of Ref.~\cite{Gabriel:2006ns} contains a charged Higgs
with partial widths to leptons and LHC production cross section
identical to those in our model.  The charged Higgs in the $Z_2$ model
differs from ours in that it can also decay to $W^+ \sigma$, where the
neutral scalar $\sigma$ is extremely light due to the spontaneous
breaking of the $Z_2$ symmetry.  This competing mode dominates unless
$H^+$ is not much heavier than $M_W$ and the neutrino Yukawa couplings
are $\mathcal{O}(1)$~\cite{Gabriel:2008es} (this parameter region is
forbidden by standard big-bang nucleosynthesis, but the $Z_2$ model
already requires nonstandard cosmology due to the very light scalar
$\sigma$).  For this parameter range, then, our results should carry
over directly.  For smaller Yukawa couplings or a heavier $H^+$, the
decays to leptons used in our analysis are suppressed, resulting in a
smaller signal on top of the same background.

Second, neutrino masses of Majorana type can be generated by the
so-called Type II seesaw mechanism~\cite{type2seesaw}, in which an
SU(2)-triplet Higgs field $X \equiv (\chi^{++}, \chi^+, \chi^0)^T$
with very small vev is coupled to a pair of SM lepton doublets.  LHC
phenomenology for this Higgs-triplet model was studied in
Ref.~\cite{Perez:2008ha}, which considered signatures from
$\chi^{++}\chi^{--}$ and $\chi^{++}\chi^-$ (and the conjugate process)
at the LHC.  While the decay branching fractions of $\chi^+$ in this
model are identical to those of the charged Higgs in our Higgs-doublet
model, the LHC production cross section for $\chi^+ \chi^-$ in the
triplet model is about 2.7 times smaller than for $H^+H^-$ in the
doublet model~\cite{Davidson:2009ha}, due to the different isospin of
$\chi^+$ which modifies its coupling to the $Z$ boson.  The signals
studied here would thus have a S/B of less than 10\% for most
channels, potentially leading to problems with background systematics.
For sufficiently high charged Higgs mass, though, the $\mu^\pm e^\mp$
channel would still have a decent S/B (33\% for $M_{H^+}=300$~GeV and
a degenerate neutrino spectrum); the reduced cross section in the
triplet model would however require an integrated luminosity close to
300 fb$^{-1}$ for discovery.  In any case, searches for the
doubly-charged scalar would yield an earlier discovery of the triplet
model.

\begin{acknowledgments}
This work was supported by the Natural Sciences and Engineering
Research Council of Canada.
\end{acknowledgments}


\end{document}